\documentclass[aps,prl,twocolumn,superscriptaddress]{revtex4-2}

\usepackage{amsmath,amssymb,amsfonts}
\usepackage{graphicx}
\usepackage{xcolor}
\usepackage{placeins}
\usepackage{float}
\usepackage{amssymb}
\usepackage{comment}
\newcommand{\tr}{\text{tr}}
\usepackage[colorlinks=true,linkcolor=blue, citecolor=blue, urlcolor=blue, bookmarks]{hyperref}
\usepackage{braket}
\usepackage{bm}
\usepackage{dsfont}

\usepackage{bbold}

\begin{document}

\title{Asymmetry dynamics and nonequilibrium symmetry-breaking phase transitions}

\author{Liv Hammer}
\affiliation{Centre for Fluid and Complex Systems, Coventry University, Coventry, CV1 2TT, United Kingdom}

\author{Colin Rylands}
\affiliation{Centre for Fluid and Complex Systems, Coventry University, Coventry, CV1 2TT, United Kingdom}

\author{Federico Carollo}
\affiliation{Dipartimento di Fisica, Sapienza Università di Roma, Piazzale Aldo Moro 5, 00185 Rome, Italy}

\date{\today}
\begin{abstract}
In classical settings, the Mpemba effect occurs when a hotter system cools faster than an initially colder one. In quantum systems, this effect can be reinterpreted exploiting the concept of symmetries, with the asymmetry of a subsystem 
playing the role of temperature. A quantum Mpemba effect arises when a more asymmetric state restores the symmetry faster than a less asymmetric one. Previous work mainly focuses on closed systems characterized by thermal equilibration and Hamiltonian symmetries. In this paper, we analyze the dynamics of asymmetry in an open quantum many-body system featuring symmetry breaking and uncover dynamical behavior that appears to be unique to these settings. In the symmetric phase, we demonstrate the existence of a quantum Mpemba effect, which  emerges as a direct consequence of a non-monotonic evolution of the asymmetry. In the broken-symmetry phase, we analyze the imbalance between the system's ability to increase or to decrease its asymmetry. Our results extend the notion of quantum Mpemba effects to open quantum many-body systems exhibiting symmetry-breaking phase transitions and establish them as a platform for observing and controlling anomalous relaxation phenomena.  
\end{abstract}

\maketitle

\begin{figure}[t]
    \centering
    \includegraphics[width=1.0\linewidth]{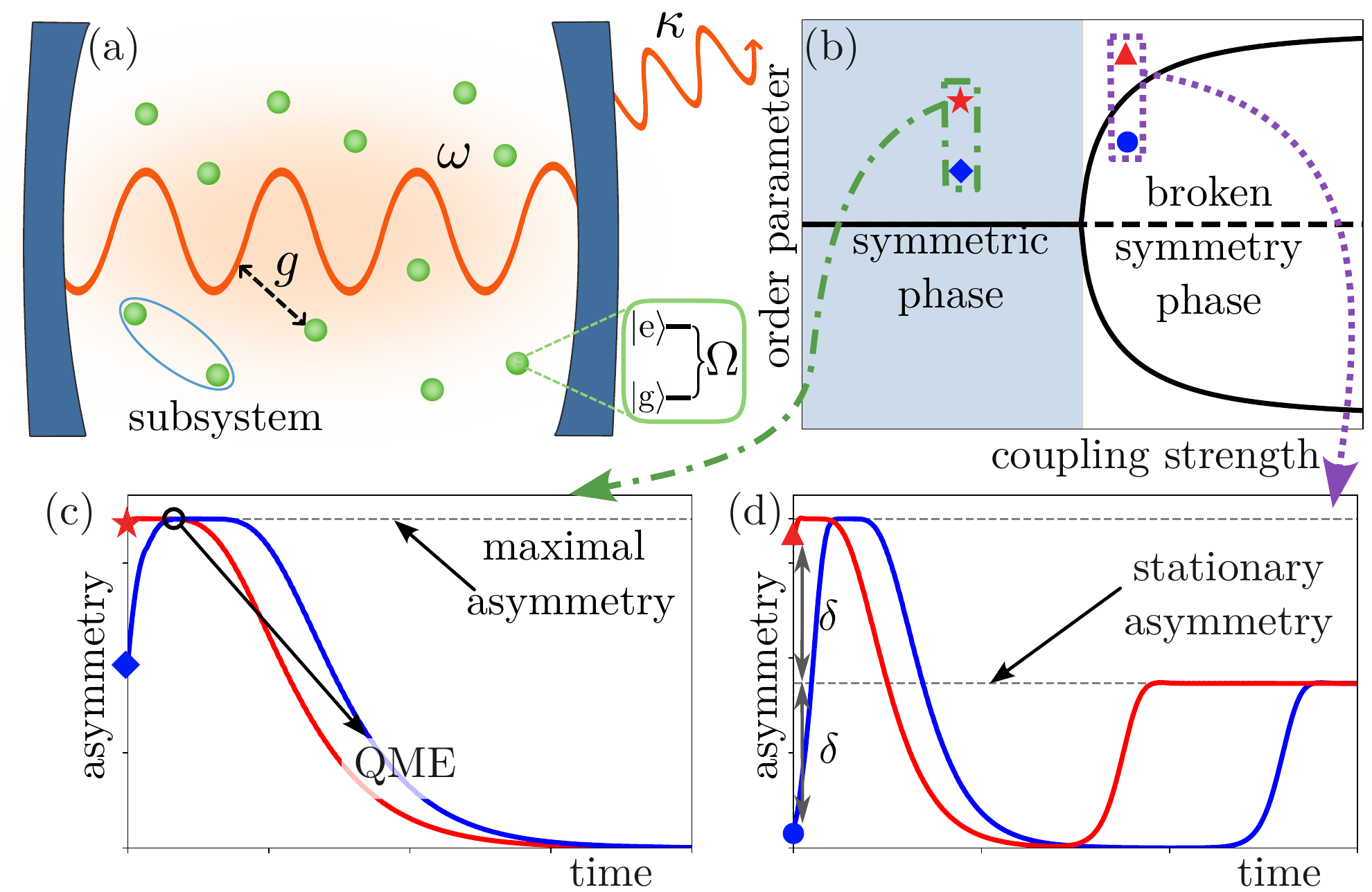}
    \caption{{\bf Open Dicke model and asymmetry dynamics.} (a) Open Dicke model, consisting of $N$ two-level atoms (energy splitting $\Omega$)  interacting with the electromagnetic field (coupling constant $g$) inside a cavity. The cavity is subject to excitation losses (rate $\kappa$). An example of a subsystem made of two atoms is highlighted. (b) The system possesses a $\mathbb Z_2$ symmetry, which is spontaneously broken for sufficiently strong coupling $g$. The symbols show  the initial conditions for the curves in (c) and (d). (c) In the symmetric phase, we observe quantum Mpemba effect (QME). A more asymmetric state (star) restores the symmetry faster than a less asymmetric state (diamond). The asymmetry dynamics is non-monotonic and fully breaks the symmetry before restoring it. (d) In the broken-symmetry phase, we observe an imbalance between the system capability to increase  asymmetry (curve originating from the circle) and to decrease it (curve originating from the triangle) by the same amount $\delta$. The curves in panels (c) and (d) are cuts indicated by the horizontal lines in  Fig.~\ref{Fig2}. }
    \label{Fig1}
\end{figure}

Far-from-equilibrium systems, be they classical or quantum, closed or open, often find unconventional routes back to equilibrium. This manifests in exotic phenomena like the Mpemba effect~\cite{Mpemba_1969_Cool}. Originally observed in classical thermal quenches, this effect entails a faster cooling for systems which are initially hotter~\cite{Teza_2026_Speedups}.  Another remarkable property of thermal quenches is the intrinsic imbalance between cooling and heating. A system relaxes quicker by heating rather than by cooling, when starting from states which are thermodynamically equidistant from the equilibrium one~\cite{Lapolla_2020_Faster,Dieball_2026_Thermal}. From a physical viewpoint, this means that increasing energy fluctuations is more efficient than decreasing them. 

Nowadays, there is significant interest in exploring analogous phenomena in quantum systems. For example, quantum Mpemba effects have been investigated in numerous contexts~\cite{Carollo_2019_Exponentially,Chatterjee_2023_Quantum,Chatterjee_2024_Multiple,Nava_2024_Mpemba,Ares_2025_Quantum,Liu_2024_Symmetry,Turkeshi_2025_Quantum,Summer_2026_Resource,Beato_2026_Relaxation,Yamashika_2026_Quantum}, and recently also observed on real  quantum devices \cite{Joshi_2024_Observing,Aharony_2024_Inverse,Zhang_2025_Observation,Xu_2025_Observation,Chatterjee_2025_Direct,Schnepper_2025_Experimental,Xia_2026_Observation}. One line of inquiry is rooted in the concept of symmetries in quantum systems. It establishes an analogy between temperature in thermal quenches and the {\it degree of asymmetry} in the relaxation dynamics of a suitable subsystem \cite{Ares_2023_Entanglement,Yamashika_2024_Entanglement,Rylands_2024_Microscopic,Ferro_2024_Nonequilibrium,Chalas_2024_Multiple,Summer_2026_Resource,Ares_2025_Simpler}. 
The typical protocol requires the initialization of the system in a state that breaks a global symmetry of the dynamics and tracks its restoration, or lack thereof, as stationarity is approached. The quantum Mpemba effect occurs when an initially more asymmetric state (hotter in the language of thermal quenches) restores the symmetry faster than an initially more symmetric one. Such a quantum Mpemba effect has been predominantly investigated in closed quantum quenches, where, due to the emergence of thermal equilibrium at long times, symmetry restoration is expected in all but a few special cases~\cite{Ares_2023_Lack,Rylands_2024_Dynamical}. While some studies have examined the role of dephasing and dissipation~\cite{Caceffo_2024_Entangled,Ares_2025_Quantum_b,Digiulio_2025_Measurement,Russotto_2025_Dynamics}, the dynamics of asymmetry in the presence of nonequilibrium symmetry breaking and stationary asymmetric states, as they can occur in open quantum systems, remains unexplored. 

In this work, we shed light on how these phenomena impact on symmetry restoration and on the asymmetry dynamics. We focus on an exemplary open quantum many-body system, the open Dicke model [see sketch in Fig.~\ref{Fig1}(a)], which can host a $\mathbb{Z}_2$ symmetry breaking in the stationary state \cite{Dimer_2007_Proposed,Baumann_2011_Exploring,Baumann_2010_Dicke,Mivehvar_2021_Cavity,Kirton_2019_Introduction,Kirton_2017_Suppressing,Carollo_2021_Exactness,Boneberg_2022_Quantum}, as illustrated in Fig.~\ref{Fig1}(b). 
We consider the asymmetry of finite subsystems and quantify it using the relative entropy of asymmetry (REA)~\cite{Gour_2009_Measuring,Marvian_2014_Extending,Casini_2019_Entanglement,Casini_2021_Entropic,Ares_2023_Entanglement}. 
In the symmetric phase, the REA can evolve in a non-monotonic way: to restore the symmetry, the subsystem may have to maximally break it first, which paves the way for the emergence of quantum Mpemba effect [cf.~Fig.~\ref{Fig1}(c)]. 
A similar behavior is also discussed for an emergent $U(1)$ symmetry defined by the  stationary state.
Within the broken-symmetry phase the REA approaches a finite value at long times and  also displays non-monotonic evolution. This includes periods of maximal symmetry breaking and/or of almost perfect symmetry restoration before reaching the stationary value [see Fig.~\ref{Fig1}(d)]. In this case, we uncover an effect analogous to the heating/cooling imbalance in thermal quenches~\cite{Lapolla_2020_Faster}. In our  setting, it entails an imbalance in the ability of the system to increase or decrease its asymmetry. We find that increasing the asymmetry can be either more or less efficient than decreasing it, depending on the initial system state. 
 
Our work extends quantum Mpemba effects---and related anomalous relaxation phenomena---to subsystems of open quantum many-body systems featuring phase transitions where it uncovers novel dynamical behavior. Our results establish a framework for investigating  nonequilibrium asymmetry dynamics in driven-dissipative quantum systems accessible with state-of-the-art  experiments \cite{Ritsch_2013_Cold,Mivehvar_2021_Cavity,Schlawin_2022_Cavity}. \\

\noindent {\bf System and symmetry operators.---} The open quantum Dicke model consists of $N$ two-level atoms, with energy splitting $\Omega$, which are placed inside a lossy single-mode cavity~\cite{Kirton_2019_Introduction}, see a sketch in Fig.~\ref{Fig1}(a).  Its Hamiltonian takes the form 
\begin{equation}\label{Hamiltonian}
    H = \omega a^\dagger a + \Omega \sum_{j=1}^N \sigma_z^{(j)} + \frac{g}{\sqrt{N}}\sum_{j=1}^N\sigma_x^{(j)} (a^\dagger + a)\, ,
\end{equation}
where $a,a^\dagger$ are bosonic annihilation and creation operators modelling the cavity light field (characteristic frequency $\omega$).  The operators $\sigma^{(j)}_{\rm a}$ (${\rm a} = x,y,z$) are Pauli matrices and the superscript denotes the atom they correspond to. The interaction term (coupling strength $g$) describes a collective coupling between the atoms and  the light field.  To ensure a meaningful thermodynamic limit, this term is renormalized by  a factor $1/\sqrt{N}$ \cite{Kirton_2019_Introduction, Carollo_2021_Exactness}. 

Due to the presence of cavity losses (rate $\kappa$), the evolution of the state of the system, $\rho (t)$, is governed by the quantum master equation $\dot \rho (t)= \mathcal{L}[\rho(t)]$, with the Lindblad generator
\begin{equation} \label{Lindblad Cavity Loss}
\mathcal{L}[\rho]= -i[H, \rho] + \kappa \left( a \rho a^\dagger -\frac{1}{2}\left\{a^\dagger a, \rho\right\} \right).
\end{equation}
Such a generator features a weak $\mathbb{Z}_2$ symmetry \cite{Buca_2012_Note,Albert_2014_Symmetries,Minganti_2018_Spectral} \footnote{Note that the model also features a strong symmetry, the total angular momentum. We therefore restrict to the fully symmetric sector to eliminate the reducibility associated with this strong symmetry.}. That is, the Lindblad generator obeys $\mathcal{L}[P \rho P] = P\mathcal{L}[\rho]P$, with $P$ being the parity operator 
\begin{equation}\label{parity op full}
    P = e^{i\pi a^\dagger a} \prod_{j=1}^N \sigma_z^{(j)}\, .
\end{equation}
This implies that a symmetric initial state of the full system, $P\rho(0) P=\rho(0)$, remains symmetric in time. It further entails that the asymmetry of the full system cannot increase during the dynamics \cite{Gour_2009_Measuring,Summer_2026_Resource}. 

To investigate whether and how the system can dynamically restore the symmetry, we consider a generic subsystem $A$ made of a finite number of atoms $\ell$ [see Fig.~\ref{Fig1}(a)]. The reduced state $\rho_A(t)$ of the subsystem is obtained by tracing over the remainder of the system. That is, $\rho_A(t)={\tr}_{{A}'} [\rho(t)]$, with ${A}'$ denoting the complement of $A$. When focussing on subsystems of atoms, we can also define the reduced symmetry operators
\begin{equation}\label{eq"subsytem_parity]}
    P_{A} = \bigotimes_{j=1}^\ell \sigma_z\, , 
\end{equation}
which act solely on the $\ell$ atoms forming the subsystem. Unlike the parity operator $P$, the operator $P_A$ does not encode the symmetry of the dynamics. It rather represents its {\it local} action on $A$. As such, an initial reduced state $\rho_A(0)$ which is (weakly) symmetric under the action of $P_A$, i.e., $P_A\rho_A(0)P_A=\rho_A(0)$, does not necessarily remain symmetric, since fluctuations of the  symmetry can enhance the local asymmetry. 

In the following, we  focus on the symmetry properties of the state $\rho_A(t)$ and investigate their dynamics, starting from initial pure states characterized by the light field in the vacuum  and the atoms in a translation-invariant product state. \\

\begin{figure*}
    \centering
    \includegraphics[width=\textwidth]{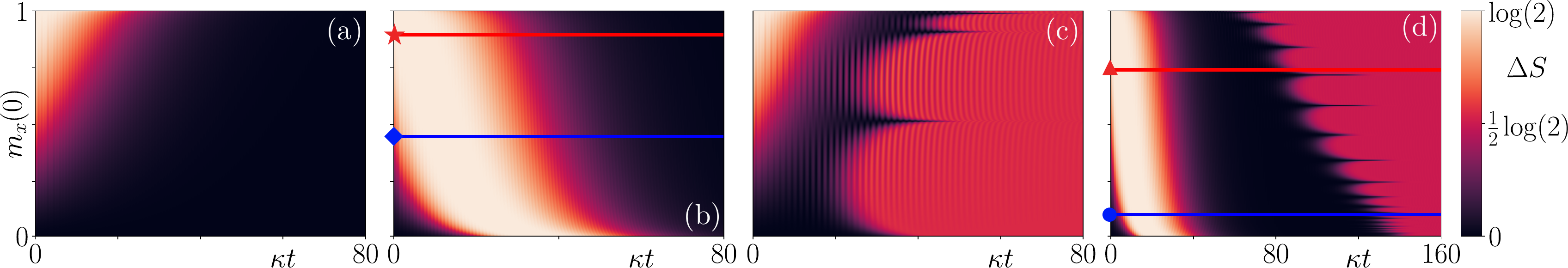}
    \caption{{\bf Asymmetry dynamics in the open quantum Dicke model.} (a) Density plot showing the time-evolved REA, $\Delta S(t,\ell)$, in the symmetric phase, for pure initial states (see end of the caption) with different values of $m_x(0)$ and with  negative $m_z(0)$. The asymmetry dynamics shows an overall monotonic behavior and small oscillations around it.  (b) Same as in (a) but for initial states with positive $m_z(0)$. The asymmetry dynamics is non-monotonic and $\Delta S(t,\ell)$ reaches its maximum value before decreasing again and vanishing at stationarity. The horizontal lines correspond to the curves in Fig.~\ref{Fig1}(c).  (c)  Time-evolved REA $\Delta S(t,\ell)$, in the symmetry-broken phase, for initial pure states with different values of $m_x(0)$ and negative $m_z(0)$. (d) Same as in (c) but for initial states with positive $m_z(0)$. The horizontal lines correspond to the curves in Fig.~\ref{Fig1}(d). All plots are for $\ell=3$, $\Omega=2\kappa$ and $\omega=1.5\kappa$. We take $g= 0.9 g_{\rm c}$ for panels (a,b) and $g/\kappa$ is such that $\Delta S(\infty,3)=\log (2)/2$ ($g/\kappa\approx 1.345>g_{\rm c}/\kappa$) for panels (c,d). The initial states are such that $m_y(0)=\alpha=0$ and $m_z(0)=-\sqrt{1-m_x^2(0)}$ for (a,c) and $m_z(0)=\sqrt{1-m_x^2(0)}$ for (b,d).}
    \label{Fig2}
\end{figure*}

\noindent {\bf Subsystem dynamics and symmetry breaking.---} In the thermodynamic limit, the dynamics of the system is captured exactly by a mean-field theory \cite{Kirton_2017_Suppressing,Kirton_2019_Introduction,Carollo_2021_Exactness,Carollo_2024_Applicability}. This provides an efficient way to explore the behavior of local observables and order paremeters. For the atom subsystem, the time-evolved reduced state reads \cite{Carollo_2021_Exactness,Mattes_2025_Long-Range}
\begin{equation}
\lim_{N\to\infty}\rho_A(t)=\bigotimes_{j=1}^\ell \rho_{\rm at}(t)\, ,
\label{}
\end{equation}
with the single-atom density matrix being 
\begin{equation}
\rho_{\rm at}(t)=\frac{1}{2}\left[\mathbb{1} +m_x(t)\sigma_x+m_y(t)\sigma_y+m_z(t)\sigma_z\right]\, .
\label{}
\end{equation}
Here, the variables $m_x,m_y,m_z$ represent  the expectation values of the Pauli matrices and evolve through the mean-field equations \cite{Kirton_2017_Suppressing,Kirton_2019_Introduction,Carollo_2021_Exactness}
\begin{equation} 
\begin{split}
    \dot m_{x} &= 2\Omega m_y, \quad\dot m_{y} = -2\Omega m_x + 2g(\alpha + \alpha^{*})m_z \, ,\\
    \dot m_{z} &= -2g(\alpha + \alpha^{*})m_y~,~
    \dot \alpha = -\big(i\omega + \frac{\kappa}{2}\big)\alpha -ig m_x\, ,\label{mean field} 
    \end{split}
\end{equation}
where $\alpha$ is the rescaled expectation value $\alpha=\lim_{N\to\infty}\langle a\rangle /\sqrt{N}$ \cite{Kirton_2019_Introduction,Carollo_2021_Exactness,Boneberg_2022_Quantum}. 
The subsystem state thus remains in product form throughout the time evolution and the dynamics further preserves its purity. This can be seen from the fact that $R=m_x^2(t)+m_y^2(t)+m_z^2(t)=1$ is a constant of motion. 

By setting the right hand side of Eqs.~\eqref{mean field} to zero, one can find the fixed points of the dynamics and study their stability \cite{Kirton_2019_Introduction}. The system exhibits a phase transition at a critical value of the coupling strength $g_{\rm c} = \sqrt{\Omega (4\omega^2 + \kappa^2)/(8\omega)}$,
that separates two nonequilibrium phases. 
When $g < g_{\rm c} $ there exists a unique (stable) stationary state characterized by $m_z=-1$ and $\alpha=0$.  When $g > g_{\rm c} $ there are two (stable) stationary states. They are characterized by $m_z=-(g_{\rm c}/g)^2$, $m_x=\pm \sqrt{1-m_z^2}$, along with $\alpha = \frac{\pm g(\omega + i\kappa/2)}{\omega^2 + \kappa^2/4}m_x $
\cite{Boneberg_2022_Quantum}. These states break the symmetry of the system since they assume a nonzero expectation  of both $m_x$ and $\alpha$. 
 \\

\noindent {\bf Relative entropy of asymmetry.---} An effective way to characterize the degree of asymmetry of a quantum state is by comparing  it with its symmetrized counterpart. Given a time-evolving  reduced state $\rho_A(t)$, we define its symmetrization through the operator $P_A$ as
\begin{equation}
\overline\rho_{A}(t)= \frac{1}{2}[\rho_A(t) + P_A\rho_A (t)P_A ] \, .
\end{equation}
We then compute the distinguishability of $\rho_A(t)$ and $\overline{\rho}_{A}(t)$ through the so-called relative entropy of asymmetry  \cite{Gour_2009_Measuring,Marvian_2014_Extending} 
\begin{equation}\label{rea}
    \Delta S(t,\ell) = \tr[\rho_A(t) (\log \rho_A(t) - \log \overline \rho_A(t))]\, ,
\end{equation}
also known as  entanglement asymmetry in closed systems \cite{Ares_2023_Entanglement} or entropic order parameter~\cite{Casini_2019_Entanglement,Casini_2021_Entropic}. 
This quantity is non-negative, $\Delta S(t,\ell) \geq 0$, and only vanishes when the subsystem state is symmetric, i.e., $\rho_A(t) = \overline{\rho}_A(t)$. For the considered symmetry operator $P_A$, it is bounded from above by $\log (2)$~\cite{Ferro_2024_Nonequilibrium,Capizzi_2024_Universal}. Exploiting the fact that the reduced subsystem state $\rho_A(t)$ is a pure product state at all times \cite{Carollo_2021_Exactness,Mattes_2025_Long-Range}, we can derive an explicit formula for the REA of a generic subsystem \footnote[11]{See supplemental material, which contains references~\cite{NielsenChuang00,Xavier_2018_Equipartition,Goldstein_2018_Symmetry}, for details on: (i) Properties of REA; (ii) Derivation of expressions for the REA;  (iii) Additional plots.}. \vphantom{\cite{NielsenChuang00,Xavier_2018_Equipartition,Goldstein_2018_Symmetry}} This is given by 
\begin{equation}\label{eq:Z2_asymm_main}
    \Delta S(t,\ell)=-\sum_{\sigma=\pm}p_\sigma(t,\ell)\log[p_\sigma(t,\ell)]
\end{equation}
where $p_\pm(t,\ell)=(1\pm [m_z(t)]^\ell)/2$ and $m_z(t)$ obeys Eqs.~\eqref{mean field}. From this expression, we note that the REA vanishes when $m_z=\pm 1$ and is maximized by $m_z=0$. 

Within this framework, the quantum Mpemba effect manifests as follows~\cite{Ares_2023_Entanglement,Rylands_2024_Microscopic}. Consider two different initial states of an $\ell$-atom subsystem, with respective REAs, $\Delta S_1(0,\ell)$ and $\Delta S_2(0,\ell)$, such that $\Delta S_1(0,\ell)>\Delta S_2(0,\ell)$. One observes a quantum Mpemba effect if there exists a $\tau$ such that $\Delta S_1(t,\ell)<\Delta S_2(t,\ell)$, $\forall t>\tau$ [see a representative case in Fig.~\ref{Fig1}(c)]. 
 \\

\noindent {\bf Dynamics in the symmetric phase.---} For $g<g_{\rm c}$ the stationary state of the subsystem is $\mathbb{Z}_2$ symmetric so that its REA vanishes, irrespective of the considered initial state. During the transient dynamics, however, $\Delta S(t,\ell)$ is strongly dependent on the initial condition. In Fig.~\ref{Fig2}(a,b) we show the behavior of $\Delta S(t,\ell)$, as a function of time, for initial states with different values of $m_x(0)$ and (a) $m_z(0)<0$, (b) $m_z(0)>0$. For $m_z(0)<0$ the REA has an overall monotonic decrease toward the stationary value and only displays small oscillations around this decay. In this regime, we observe that the more the symmetry is broken by the initial state, the longer the time it takes for it  to be fully restored. The situation is drastically different when $m_z(0)>0$, which is shown in Fig.~\ref{Fig2}(b). In this case, the REA  evolves non-monotonically, first increasing and assuming the maximum value before eventually decaying to zero as the symmetry is fully restored. The plot in Fig.~\ref{Fig2}(b) shows that this non-monotonic dynamics is such that more asymmetric initial states restore the symmetry faster than more symmetric ones [see also curves in Fig.~\ref{Fig1}(c)], indicating the emergence of a quantum Mpemba effect. 

Interestingly, even symmetric subsystem initial states can develop a non-zero asymmetry over time. This is not the case for the states considered in Fig.~\ref{Fig2}(a,b). There, indeed, the symmetric states are those with $m_x(0)=m_y(0)=0$ and $m_z(0)=\mp 1$. Given that we further assumed $\alpha(0)=0$, these states are the (stable and unstable) fixed points of Eqs.~\eqref{mean field}. However, one can for instance take an initial symmetric subsystem state [$m_x(0)=m_y(0)=0$] and consider $m_z(0),\alpha\neq 0$. This will develop a finite value of $m_x(t)$ and $m_y(t)$  [cf.~Eqs.~\eqref{mean field}], yielding a non-zero REA at finite times. 
\\

\noindent {\bf Dynamics in the broken-symmetry phase.---}
In the phase in which the system spontaneously breaks the $\mathbb{Z}_2$ symmetry, the REA converges to a finite  stationary value. The latter solely depends on the value of the dynamical parameters. The transient dynamics is instead strongly dependent on the initial state. In Fig.~\ref{Fig2}(c,d) we see that for both $m_z(0)<0$ (c) and $m_z(0)>0$ (d) the evolution of the REA is non-monotonic. It goes through a regime of almost perfect symmetry restoration prior to eventually attaining the stationary value. The extent of time during which the state of the system stays close to perfect symmetry restoration depends on how far $g$ is from the critical value, increasing the closer $g$ is to $g_{\rm c}$. 

The existence of two possible stationary states creates a richer dynamical structure. Whether the system relaxes to the upper or lower branch of the phase diagram depicted in Fig.~\ref{Fig1}(b) depends non-trivially on the initial state \cite{Note11}. As a result, only slightly different initial conditions can approach opposite stationary states. This can lead to differences in relaxation times, in fact, it appears to cause the cusp-like behavior shown by the REA in Fig.~\ref{Fig2}(c,d). In~Fig.~\ref{Fig1}(d), we compare the evolution of the REA for two initial states  with $\Delta S(0,\ell)=\Delta S(\infty,\ell)\pm \delta$. The state which initially is less asymmetric demonstrates a longer stationary state relaxation time. The same overall trend is shown  in~Fig.~\ref{Fig2}(d), where the value of $g$ is chosen such that the stationary value of the REA is given by $\Delta S(\infty,\ell)=\log(2)/2$. Comparing, for instance, a state with $\Delta S(0,\ell)\to\log (2)$ to one with $\Delta S(0,\ell)\to 0$, we see that the former will reach stationarity first. This indicates that the system can more efficiently decrease asymmetry than increase it, which is reminiscent of the cooling/heating imbalance in thermal quenches~\cite{Lapolla_2020_Faster,Dieball_2026_Thermal}. Unlike thermal quenches, however, the imbalance is not always in the same direction and it is also possible to observe the more symmetric states relaxing faster, as for instance clearly displayed in Fig.~\ref{Fig2}(c). \\

\begin{figure}[t]
    \centering
\includegraphics[width=1.0\linewidth]{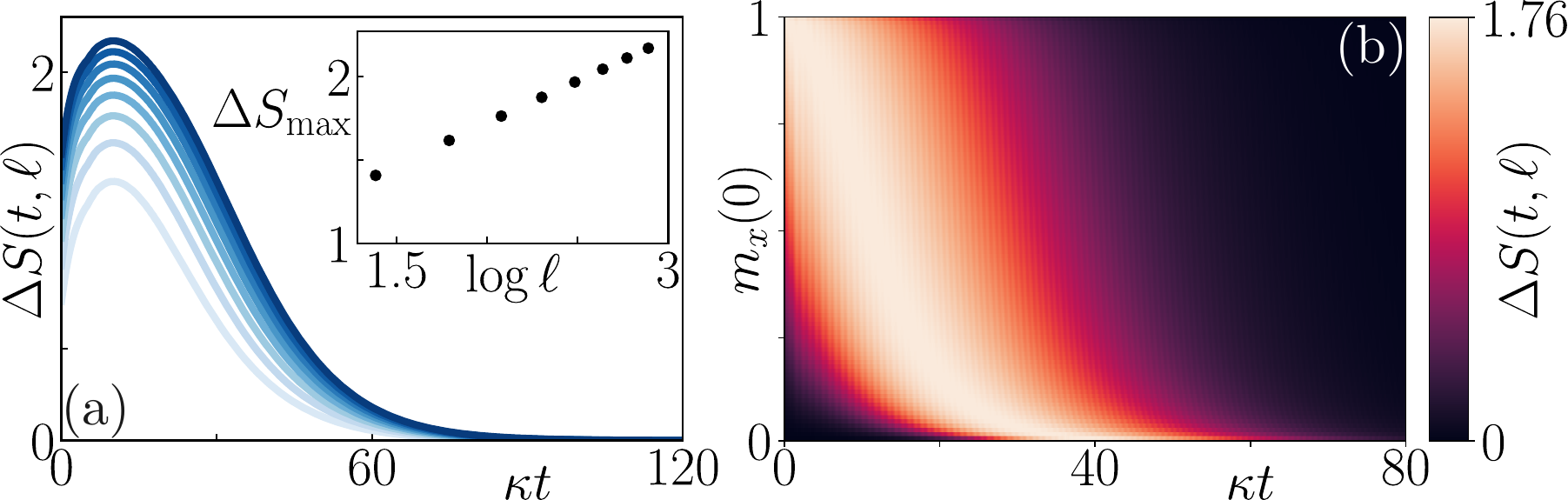}
    \caption{{\bf  Dynamical emergence of $U(1)$ symmetry.} (a) Time-evolved REA, $\Delta S(t,\ell)$, for the emergent $U(1)$ symmetry in the symmetric phase. The main plot shows results for $\ell=4,6,\dots, 18$, while the inset reports the dependence of the maximum of the REA, $\Delta S (t, \ell)$, over time for different $\ell$. (b) Density plot of the REA for the emergent $U(1)$ symmetry as a function of time, in the symmetric phase, $g/g_{\rm c}=0.9$. We consider initial pure states with different values of $m_x(0)$, $m_y(0)=\alpha=0$ and positive $m_z(0)$. Here, we have $\ell=8$ and all other parameters are as in Fig.~\ref{Fig2}.}
    \label{Fig3}
\end{figure}

\noindent {\bf Emergent symmetry.---} The REA provides a way to track the dynamical restoration, or breaking, of a symmetry at the level of subsystems. It is not restricted to explicit symmetries of the dynamical generators and can also be used to explore emergent ones~\cite{Ares_2025_Entanglement,Klobas_2025_Translation,Yu_2025_Symmetry}. For instance, in the symmetric phase, the subsystem stationary state exhibits an emergent $U(1)$ symmetry implemented by the unitary  $W(\lambda)=\bigotimes_{j=1}^\ell e^{i\lambda \sigma_z}$ [it is indeed an eigenstate of $W(\lambda)$]. An analytic expression for the REA with respect to this $U(1)$ symmetry is derived in~\cite{Note11}. In Fig.~\ref{Fig3}, we show its time evolution for different subsystem sizes and  initial states. We observe qualitatively similar behavior to the case of the explicit $\mathbb{Z}_2$ symmetry. The subsystem evolves through a maximally asymmetric regime, where at some time $t^*$, $\Delta S(t^*,\ell)\sim \frac{1}{2}\log(\ell)$, before decaying to zero [see inset of Fig.~\ref{Fig3}(a)]. Comparing the dynamics of different initial states we see that this non-monotonic behavior can also result in a quantum Mpemba effect as clearly displayed by the density plot in Fig~\ref{Fig3}(b). The same quantity can also be investigated in the broken-symmetry  phase~\cite{Note11}.  \\

\noindent {\bf Discussion.---} We have studied the dynamics of the subsystem asymmetry, using the relative entropy of asymmetry, in the open quantum Dicke model. The system exhibits a nonequilibrium symmetry-breaking phase transition and hosts a rich structure of dynamical behavior. We have shown the possibility of quantum Mpemba effects in the symmetric phase and furthermore discussed the dynamics and the approach to stationarity in the broken-symmetry phase. 

Mpemba effects have recently been given a unified framework through the language of resource theories~\cite{Summer_2026_Resource,Aditya_2025_Mpemba}. Therein, an Mpemba effect occurs if a state dissipates a resource (asymmetry in our case or athermality in the case of thermal quenches) quicker, if it is initially more resourceful. Our results in the broken-symmetry phase can be phrased in an analogous way: we find an imbalance between the subsystem ability to dynamically generate or dissipate its resource, namely asymmetry, when subject to a same dynamics. This opens the door to exploring similar effects using other quantum resources, such as non-Gaussianity or non-stabilizerness. 

Here, we have focussed on the asymmetry of subsystems for two key reasons: $i)$ it can show richer (e.g., non-monotonic) dynamical behavior; $ii)$ in real experiments with many-body systems, it can be practically challenging to reconstruct the full state of the system, while local (order-parameter) properties may be more accessible. In the future, one could also investigate the dynamics of the asymmetry in our model, or related ones, at the level of the full quantum state. In this case, the asymmetry can only decrease yet quantum Mpemba effects could still be observed. For instance, one could numerically investigate the full state asymmetry for a finite-$N$ system, similarly to what was done for a single-body toy-model in Ref.~\cite{Summer_2026_Resource}. 
\\

\textbf{Acknowledgements.---} We acknowledge Sascha Wald for bringing Ref.~\cite{Lapolla_2020_Faster} to our  attention. \\

\textbf{Data availability.---} The data displayed in the figures is
available on Zenodo \cite{zenodo2026}.

\bibliography{references}

\newpage
\setcounter{equation}{0}
\setcounter{figure}{0}
\setcounter{table}{0}
\makeatletter
\renewcommand{\theequation}{S\arabic{equation}}
\renewcommand{\thefigure}{S\arabic{figure}}
\makeatletter

\onecolumngrid
\newpage

\setcounter{page}{1}
\begin{center}
{\Large SUPPLEMENTAL MATERIAL}
\end{center}
\begin{center}
\vspace{0.8cm}
{\Large Asymmetry dynamics and nonequilibrium symmetry-breaking phase transitions}
\end{center}
 \begin{center}
 Liv Hammer$^1$, Colin Rylands$^1$, Federico Carollo$^2$
 \end{center}
 \begin{center}
 $^1${\em Centre for Fluid and Complex Systems, Coventry University, Coventry, CV1 2TT, United Kingdom}\\
 $^2${\em Dipartimento di Fisica, Sapienza Università di Roma, Piazzale Aldo Moro 5, 00185 Rome, Italy}\\
 \end{center}
\section{Properties of the REA}
\noindent The relative entropy of asymmetry is defined as 
\begin{eqnarray}
    \Delta S(t,\ell)=\tr[\rho_A(t)\log\rho_A(t)]-\tr[\rho_A(t)\log\bar{\rho}_A(t)]
\end{eqnarray}
where $\bar\rho_A(t)$ is the symmetrized version of $\rho_A(t)$. For either of the symmetry groups we consider, $\mathbb{Z}_2$ or $U(1)$, this is defined as 
\begin{eqnarray}
    \bar\rho_A(t)=\sum_{q}\Pi_q\rho_A(t)\Pi_q\, .
\end{eqnarray}
Here, the sum is over eigenvalues $q$ of the associated group generator and $\Pi_q$ is the projector onto the eigenspace with eigenvalue $q$. The $\mathbb{Z}_2$ group is generated by $P_A$ which has two eigenvalues $\pm 1$. We denote the projectors by 
\begin{eqnarray}
    \Pi_\pm=\frac{1}{2}(1\pm P_A)~.
\end{eqnarray}
The symmetrised state is then given by 
\begin{equation}\label{eq:symm_stat_z2}
\begin{split}
    \bar\rho_A(t)&=\Pi_+\rho_A(t)\Pi_++\Pi_-\rho_A(t)\Pi_-\\
    &=\frac{1}{2}\big(\rho_A(t)+P_A\rho_A(t)P_A\big) \, ,
    \end{split}
\end{equation}
which is the expression quoted in the main text. For the $U(1)$ symmetry, the group is generated by the magnetization along $z$, $S^z_A=\sum_{j\in A}\sigma^{(j)}_z$. (Here, with a slight abuse of notation we denote $\sigma_z^{(j)}$ the Pauli matrix acting only on site $j$ but embedded in the subsystem of size $\ell$ rather than in the full system which would be the notation used in the main text.) The symmetrized state is 
\begin{eqnarray}
    \bar\rho_A(t)=\sum_{q=-\ell}^\ell \Pi_q \rho_A(t)\Pi_q \, ,
\end{eqnarray}
where $q=-\ell,\dots, \ell$ are $2\ell+1$ the eigenvalues of $S^z_A$ for $\ell$ spins and $\Pi_q$ the projectors onto the eigenspace with eigenvalue $q$. We can use the following representation of this projector
\begin{equation}\label{eq:U1_proj}
    \Pi_q = \int_{-\pi}^\pi \frac{{\rm d}\lambda}{2\pi} e^{i\lambda(S^z_A - q)},
\end{equation}
from which we find that
\begin{eqnarray}
    \bar\rho_A(t)= \int_{-\pi}^\pi \frac{{\rm d}\lambda}{2\pi}e^{i \lambda S^z_A }\rho_A(t)e^{-i\lambda S^z_A }~.
\end{eqnarray}
This expression is analogous to \eqref{eq:symm_stat_z2} but for the case of the $U(1)$ symmetry.

Using the definition of $\bar\rho_A(t)$ we can show that $\tr[\rho_A(t)\log\bar{\rho}_A(t)]=\tr[\bar\rho_A(t)\log\bar{\rho}_A(t)]$~\cite{Gour_2009_Measuring}. Consequently, the REA can be expressed as a difference of entropies
\begin{equation}
    \Delta S(t,\ell) = S (\bar \rho_{A}(t)) - S (\rho_{A}(t)),
\end{equation}
 where $S(\rho)=-\tr[\rho\log \rho]$. This can be simplified further  if the reduced state $\rho_A(t)$ is pure, in which case $ S (\rho_{A}(t))=0$. Using the fact that $\bar{\rho}_A(t)$ is symmetric by construction, and following the symmetry resolution of entanglement entropy~\cite{Goldstein_2018_Symmetry,Xavier_2018_Equipartition} we have,
 \begin{equation}
 \begin{split}
      \Delta S(t,\ell) &= S (\bar \rho_{A}(t)) \\\label{eq:symm_res}
      &= -\sum_q p_q(t,\ell)\log [p_q(t,\ell)]+\sum_q p_q(t,\ell)S(\rho_{A,q}(t))\, .
      \end{split}
 \end{equation}
Here, we have introduced 
\begin{eqnarray}\label{eq:probs}
    p_q(t,\ell)=\tr[\Pi_q\rho_A(t)]
\end{eqnarray}
which is the probability that a measurement of the symmetry generator, $P_A$ or $S^z_A$ in the state $\rho_A(t)$ returns the value $q$. In addition, 
\begin{eqnarray}
\rho_{A,q}(t)=\frac{\Pi_q \rho_A(t)\Pi_q}{p_q(t,\ell)}
\end{eqnarray}
is the state which the system ends up in, via the Born rule, after the measurement. The expression~\eqref{eq:symm_res} states that the entropy of a symmetric state splits into two parts. The first is the entropy associated to the probability distribution of the measurement outcomes and the second is the average entropy of a state after a measurement has been made. Since our reduced state, described by $\rho_A(t)$, remains pure at all times, $\rho_{A,q}(t)$ must also remain a pure state as a measurement cannot increase the entropy~\cite{NielsenChuang00}. Thus, the second term in~\eqref{eq:symm_res} must vanish and we have that 
\begin{eqnarray}\label{REA_start_point}
    \Delta S (t,\ell)= - \sum_{q} p_q(t,\ell) \log[ p_q(t,\ell)]~
\end{eqnarray}
which is valid at all times for the states we consider and for either the $\mathbb{Z}_2$ or $U(1)$ symmetry. 

\section{Derivation of REA for $\mathbb{Z}_2$ symmetry}
 In this section of the supplemental material we provide the derivation for the REA for the open quantum Dicke model, computed with respect to its $\mathbb{Z}_2$ symmetry.
For a generic subsystem $A$, with $|A| = \ell$ atoms, the parity operator takes the form $P_A = \bigotimes_{j=1}^\ell\,\sigma_z$ yielding the projectors 
\begin{equation} \label{projection op. l spins}
    \Pi_\pm = \frac{1}{2}  \Big( \mathbb{1} \pm P_A \Big ).
\end{equation}

\noindent Since all atoms evolve identically $\rho_{A}(t) = \bigotimes_{j=1}^\ell\rho_{\rm at}(t)$, we get that the probabilities~\eqref{eq:probs} are 
\begin{equation}
\begin{split}
     p_\pm (t,\ell)&= \tr[\Pi_\pm \rho_A(t)] = \frac{1}{2} \tr \Big [  \rho_A(t) \pm \bigotimes_{j=1}^\ell[\sigma_z\rho_{\rm at}(t)]\Big ] \\\label{eq:Z2_prob}
    &= \frac{1}{2} \Big ( 1 \pm [m_z(t)] ^{\ell} \Big).
\end{split}
\end{equation}

\begin{figure}[h]
    \centering
    \includegraphics[width=.65\linewidth]{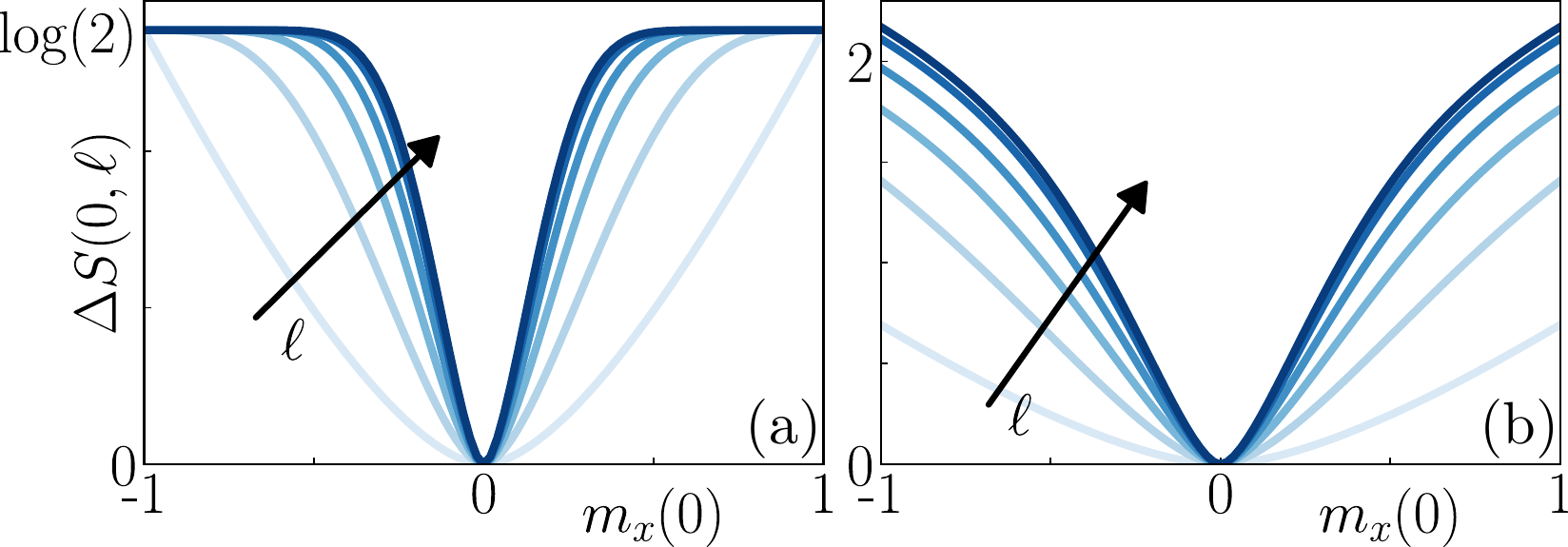}
    \caption{The initial REA, $\Delta S (0,\ell)$, as a function of $m_x(0)$, with $m_y(0)=0$, for different subsystem sizes, $\ell = 4, 8, 12, 16, 18$ (going from light to dark). (a) REA with respect to the $\mathbb{Z}_2$ symmetry.  (b) REA with respect to the $U(1)$ symmetry. }
    \label{fig:initial_asymm}
\end{figure}

\begin{figure}[h]
    \centering
    \includegraphics[width=.65\linewidth]{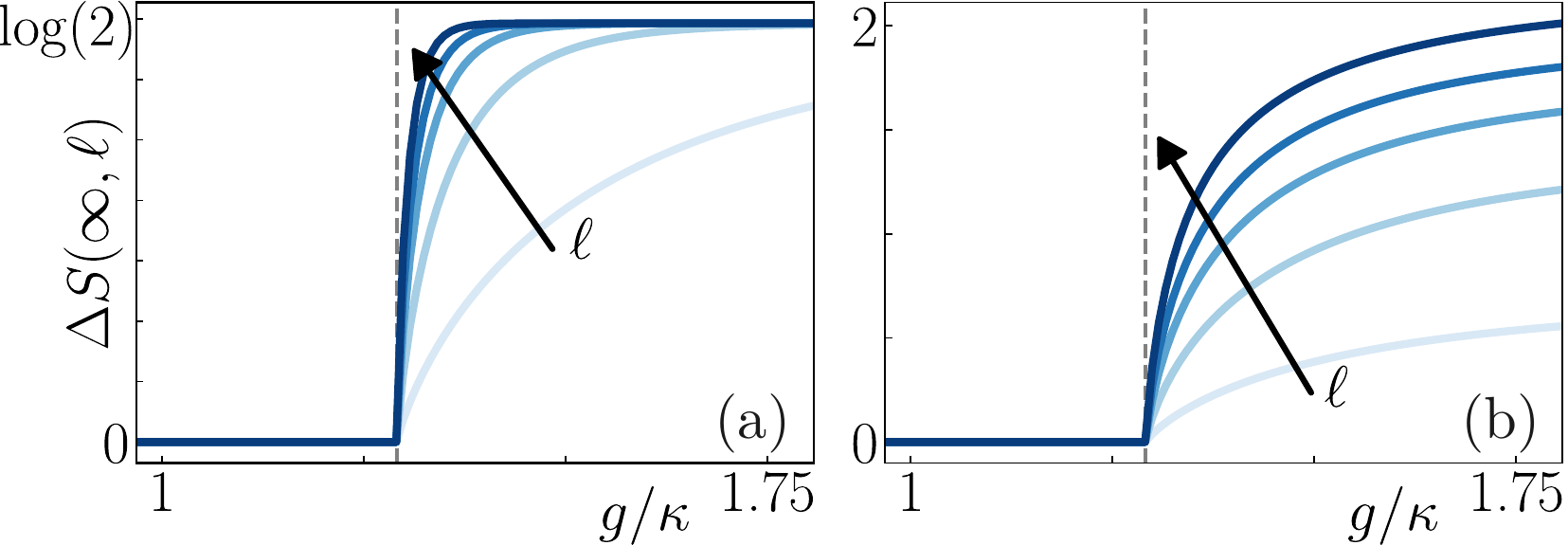}
    \caption{Stationary value of the REA, $\Delta S(\infty,\ell)$, as a function of $g/\kappa$, for different subsystem sizes, $\ell = 4, 8, 12, 16, 18$ (going from light to dark). (a) Stationary REA with respect to the $\mathbb{Z}_2$ symmetry. (b) Stationary REA with respect to the $U(1)$ symmetry. In both panels, we take the initial state $(m_x(0), m_y(0), m_z(0)) = (\sqrt{1/2}, \sqrt{1/4}, \sqrt{1/4})$, the parameters $\Omega = 2\kappa $ and $\omega = 1.5\kappa$, the dashed vertical line indicates $g_{\rm c}/\kappa$.}
    \label{fig:stationary_asymm}
\end{figure}

\noindent Inserting this  into Eq.~\eqref{REA_start_point}, we obtain the expression of the REA,
\begin{equation}\label{eq:Z2_result}
    \Delta S (t,\ell) = - \Big ( \frac{1-[m_z(t)]^\ell}{2}\Big) \log \Big ( \frac{1-[m_z(t)]^\ell}{2}\Big) - \Big ( \frac{1+[m_z(t)]^\ell}{2}\Big) \log \Big ( \frac{1+[m_z(t)]^\ell}{2}\Big) \, ,
\end{equation}
which is the one reported in the main text. In Fig.~\ref{fig:initial_asymm} (a) we plot the REA for the initial states considered in Fig.~\ref{Fig2}  using~\eqref{eq:Z2_result} for different values of $\ell$. 
In Fig.~\ref{fig:stationary_asymm}(a) we plot instead the stationary values of the REA as a function of $g$ fixing the other parameters. (Note that both stationary states in the symmetry-broken phase have the same REA.)

\section{Derivation of REA for $U(1)$ Symmetry}
\noindent Following the previous section, we extend the discussion by now considering the case of an (emergent) $U(1)$ symmetry. In our case, we have that, in the symmetric case, the stationary subsystem state is an eigenstate of the operator $S^z_A=\sum_{j\in A}\sigma_z^{(j)}$ which can be taken as an emergent $U(1)$ symmetry.

The REA is given by \eqref{REA_start_point} and the probabilities can be computed using~\eqref{eq:U1_proj},
\begin{equation}
    p_q (t,\ell)=  \int_{-\pi}^\pi \frac {{\rm d}\lambda }{2\pi}e^{-i\lambda q}\tr[e^{i\lambda S^z_A}\rho_{A} (t)].
\end{equation}
\noindent As the state is a pure product state comprised of $\ell$ identical spin it may be written as $\rho_{A} (t)= \bigotimes_{j=1}^\ell \ket{\phi}\!\bra{\phi}$, where $\ket{\phi}$ are single-atom  states. Using this, and performing the trace operation,
\begin{equation} \label{pq U1 start}
    p_q = \frac{1}{2\pi} \int_{-\pi}^\pi {\rm d}\lambda e^{-i\lambda q}[ \bra{\phi} e^{i\lambda \sigma_z} \ket{\phi}]^\ell.
\end{equation}
\noindent Rewriting the single-atom states as
\begin{align}
    \ket{\phi} &= \begin{bmatrix}
           \cos( \theta/2 ) \\
           e^{i \varphi} \sin (\theta /2)
         \end{bmatrix},
\end{align}
\noindent we can derive an expression for the term inside the brackets of the integrand in \eqref{pq U1 start}
\begin{equation}
\begin{split}
    \bra{\phi}(\cos (\lambda) + i\sin(\lambda) \sigma_z) \ket{\phi} 
    &= \begin{bmatrix}
             \cos( \theta/2 ) &
           e^{-i \varphi} \sin (\theta /2)
         \end{bmatrix}
    \begin{bmatrix}
           \cos( \lambda ) + i\sin(\lambda)& 0\\
            0 & \cos(\lambda) - i\sin(\lambda)
         \end{bmatrix}
         \begin{bmatrix}
             \cos( \theta/2 ) \\
           e^{i \varphi} \sin (\theta /2)
         \end{bmatrix} \\
         & = \begin{bmatrix}
             \cos( \theta/2 ) &
           e^{-i \varphi} \sin (\theta /2)
         \end{bmatrix} \begin{bmatrix}
             e^{i\lambda} \cos(\theta /2) \\
             e^{-i \lambda}\sin(\theta /2) 
         \end{bmatrix} = \cos^2(\theta/2)e^{i\lambda} + e^{-i\lambda} \sin^2(\theta/2) \\
         &= \frac{1}{2}(1+ \cos\theta ) e^{i\lambda} + \frac{1}{2} (1-\cos\theta)e^{-i\lambda} \\
         &=   p_+(t,1) e^{i\lambda} +  p_-(t,1)e^{-i\lambda} \label{polar to cartesian},
\end{split}
\end{equation}
\noindent where in the last line we used $m_z = \cos\theta$ and from~\eqref{eq:Z2_prob},
\begin{eqnarray}
    p_\pm(t,1)=\frac{1\pm m_z(t)}{2}~.
\end{eqnarray}
Exponentiating by $\ell$ and applying the binomial theorem 
\begin{align}
    [p_+(t,1) e^{i\lambda} + p_-(t,1) e^{-i\lambda}]^\ell  = \sum_{k=0}^\ell \binom{\ell}{k}  e^{i\lambda (\ell-2k)} [p_+(t,1)]^{\ell-k}[p_-(t,1)]^k,
\end{align}
inserting this back into \eqref{pq U1 start} we arrive at a final expression for the probabilities
\begin{equation}
\begin{split}
    p_q(t,\ell) &= \frac{1}{2\pi} \sum_{k=0}^\ell \binom{\ell}{k} \int_{-\pi}^\pi {\rm d}\lambda e^{-i\lambda q}   e^{i\lambda (\ell-2k)} [p_+(t,1)]^{\ell-k}[p_-(t,1)]^k \\
    &= \sum_{k=0}^\ell \binom{\ell}{k}  \delta_{q, \ell-2k} [p_+(t,1)]^{\ell-k}[p_-(t,1)]^k \\\label{eq:U1_result}
    &= \binom{\ell}{\frac{\ell-q}{2}} [p_+(t,1)]^{\big(\frac{\ell+q}{2}\big)}  
     [p_-(t,1)]^{\big(\frac{\ell-q}{2}\big)}\, .
\end{split}
\end{equation}
Inserting this into~\eqref{REA_start_point} we have the expression used to obtain the results in the main text. 

\begin{figure}
    \centering
    \includegraphics[width=0.65\linewidth]{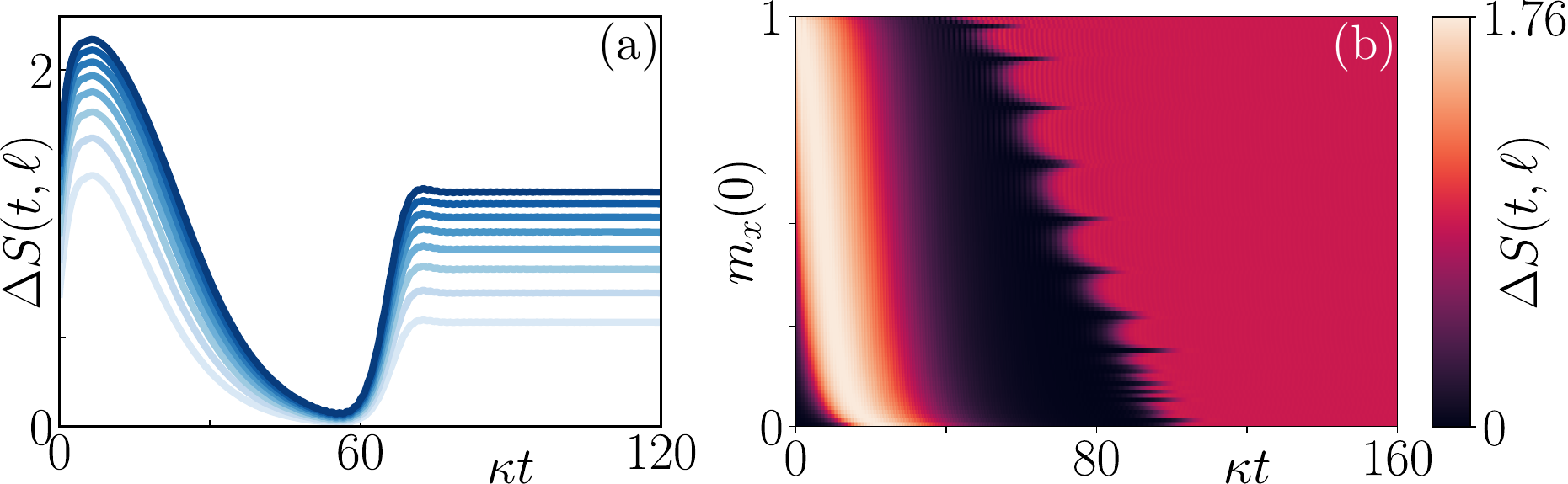}
    \caption{Dynamics of the REA, $\Delta S(t,\ell)$, with respect to the $U(1)$ symmetry in the broken-symmetry phase. We take $g/\kappa$ such that the stationary value for the REA is equal to half of the maximum REA, $\Delta S(\infty,\ell) = \frac{1}{2}\Delta S^{\max}_{\ell = 8}$ ($g/\kappa\approx 1.373$). All other parameters are as in Fig.~\ref{fig:stationary_asymm}. (a) REA as a function of time for different subsystem sizes $\ell=4, 6, \dots,18$ (going from light to dark). The initial state of the system is chosen to be $(m_x(0), m_y(0), m_z(0)) = (\sqrt{1/3}, \: 0, \sqrt{2/3})$.
    (b) Density plot over the REA as a function of time for different values of $m_x(0)$ with $m_y(0) = 0$, $m_z(0)>0$, and $\ell=8$. }
    \label{fig:U1_symm_break}
\end{figure}

In Fig.~\ref{fig:initial_asymm}(b) we plot the initial REA as a function of $m_x(0)$ using~\eqref{eq:U1_result}. 
In Fig.~\ref{fig:U1_symm_break} we plot the REA with respect to the $U(1)$ asymmetry in the symmetry-broken phase for different values of $\ell$ (a) and for different initial states (b). The REA reaches its maximum value when at some time $t^*$ defined $m_z(t^*)=0$, for which 
\begin{eqnarray}
    p_q(t^*,\ell)=\frac{1}{2^\ell} \binom{\ell}{\frac{\ell-q}{2}}~.
\end{eqnarray}
Using this we find that the maximal REA, for large subsystem sizes, is given by
\begin{eqnarray}
    \Delta S(t^*,\ell)=\frac{1}{2}\log\Big(\frac{\pi e\ell}{2} \Big)+O(1/\ell) \, .
\end{eqnarray}
Similarly, in the long time limit the stationary value of the asymmetry is
\begin{equation}
\begin{split}
    \lim_{t\to \infty} \Delta S(t,\ell)&= \frac{1}{2}\log\Big(\Big[1-\Big(\frac{g_c}{g}\Big)^{\!4}\Big] \frac{\pi e\ell}{2} \Big)+O(1/\ell)~\\
   &= \frac{1}{2}\log\Big( \frac{\pi e m_x^2\ell}{2} \Big)+O(1/\ell) \, .
   \end{split}
\end{equation}
In Fig.~\ref{fig:stationary_asymm}(b) we plot this stationary value as a function $g/\kappa$. 

\section{Stationary states in the broken symmetry phase}
\begin{figure}
    \centering
    \includegraphics[width= .65\linewidth]{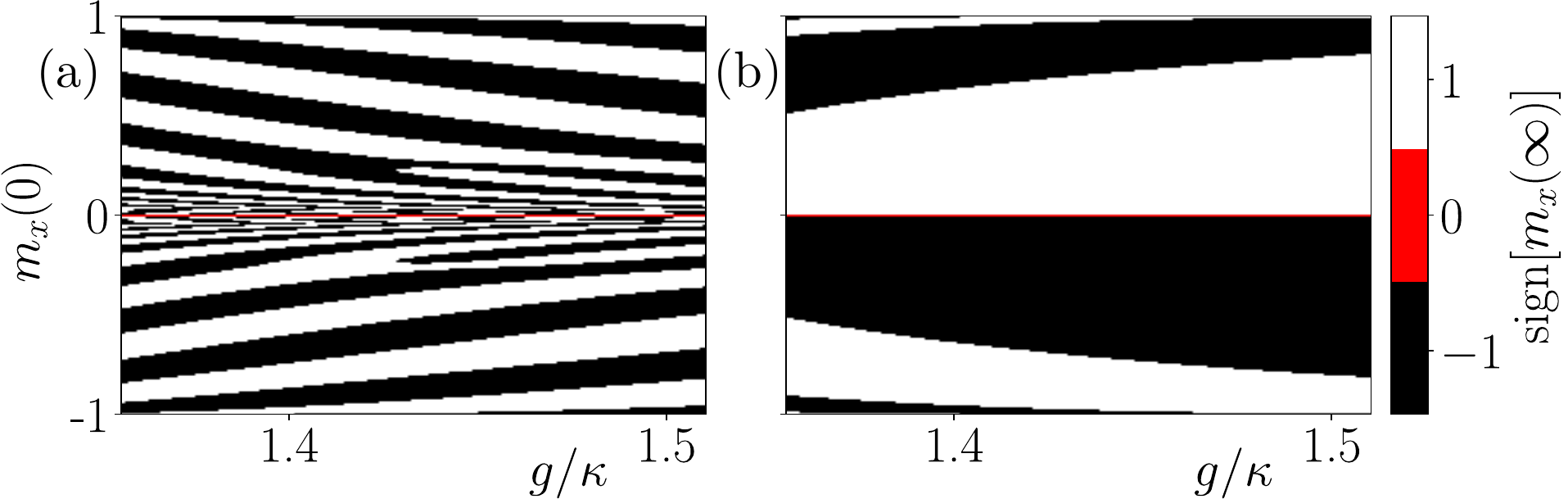}
    \caption{The value of sign$(m_x)$ in the stationary state of the broken symmetry phase as a functions of both $m_x(0)$ and $g$. The initial states are such that $m_y(0) = 0$ in both panels and (a) $m_z(0)>0$ and (b) $m_z(0)<0$. White indicates that the system  evolves to the stationary state in the upper branch of the phase diagram [Fig.~\ref{Fig1}(b) of the main text] while black indicates that it evolves to the lower branch. The red line indicates the unstable stationary state with $m_x=0$.}
    \label{fig:stationary_state_sign}
\end{figure}

\noindent In the broken symmetry phase the system has a choice of two stationary states. To investigate which initial condition ends up in which stationary state we plot the long time value of sign$(m_x)$ which is $+1$ for the upper branch of the phase diagram depicted in Fig.~\ref{Fig1}(b) and -1 for the lower branch. In Fig.~\ref{fig:stationary_state_sign} we plot this as a color map as a function of $m_x(0)$ and $g>g_c$, considering the case where $m_z(0)>0$ (a) and $m_z(0)<0$ (b). We see that in both cases the branch which is chosen depends intricately on both $m_x(0)$ and $g$. Comparing these plots with those in Fig.~\ref{Fig2}(c,d) of the main text we see that the cusp-like features appear to coincide with the interfaces between the black and white regions in Fig.~\ref{fig:stationary_state_sign}. For these particular values the states sit at the boundary between ending up on the upper or lower branch, resulting in an extended relaxation time. 

\section{Additional results on the $U(1)$ symmetry}
\begin{figure}[H]
    \centering
    \includegraphics[width=0.65\linewidth]{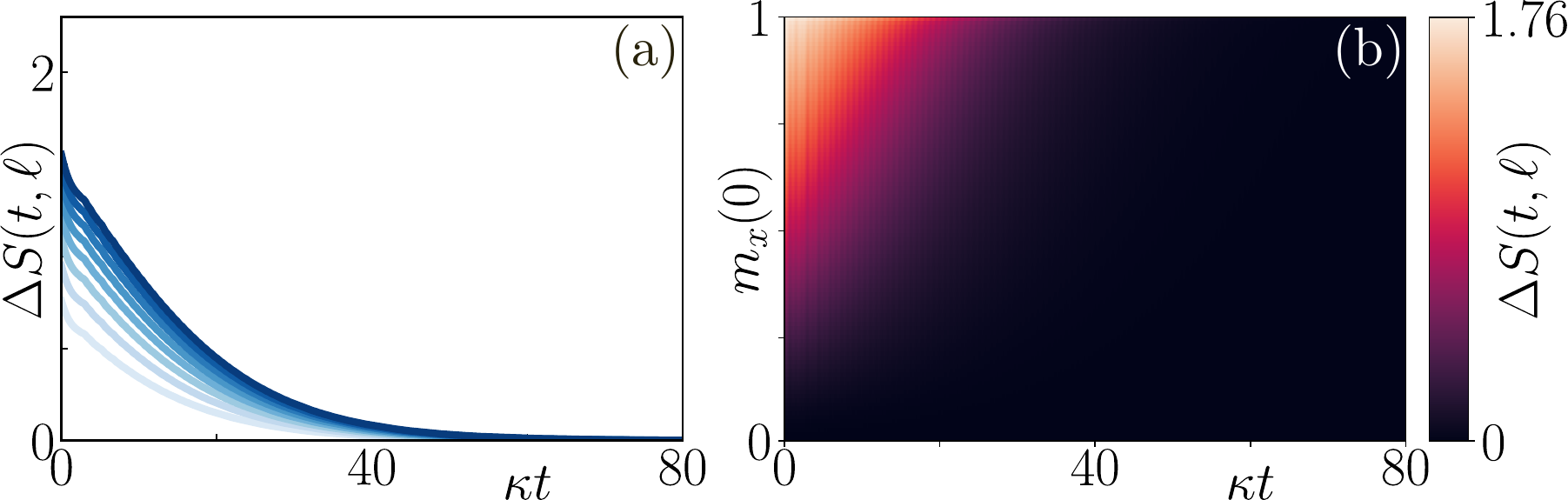}
    \caption{Dynamics of the REA, $\Delta S(t, \ell)$, with respect to the $U(1)$ symmetry in the symmetric phase. All parameters are as in Fig. ~\ref{fig:stationary_asymm} and we take $g/\kappa = 0.9 g_{\rm c}/\kappa$. (a) REA as a function of time and $m_x(0)$ for different subsystem sizes $\ell = 2, 4, 6, \dots, 18$ (going from light to dark). The initial state is as in Fig.~\ref{U1_broken_symm_zneg}. (b) Density plot of the REA as function of $t$ and $m_x(0)$ for $\ell = 8$, with $m_z(0) \leq 0$.}
    \label{U1_symmetric_zneg}
\end{figure}

\noindent The transient dynamics in the phase in which  the stationary state of the system is $\mathbb{Z}_2$ symmetric exhibit strong initial state dependence. Here, we see that a similar behavior can be observed for the case of the emergent $U(1)$ symmetry. We consider the same class of initial states. In Fig.~\ref{U1_symmetric_zneg} we see that the dynamics of the REA  with respect to $U(1)$ symmetry for initial states with $m_z(0) < 0$ show an overall monotonic decay towards stationarity, as in the case for $\mathbb{Z}_2$ symmetry shown in Fig.~\ref{Fig2}(a). In this phase, initial states with greater asymmetry have longer relaxation times than the more symmetric initial states. 

\begin{figure}[H]
    \centering
    \includegraphics[width=0.65\linewidth]{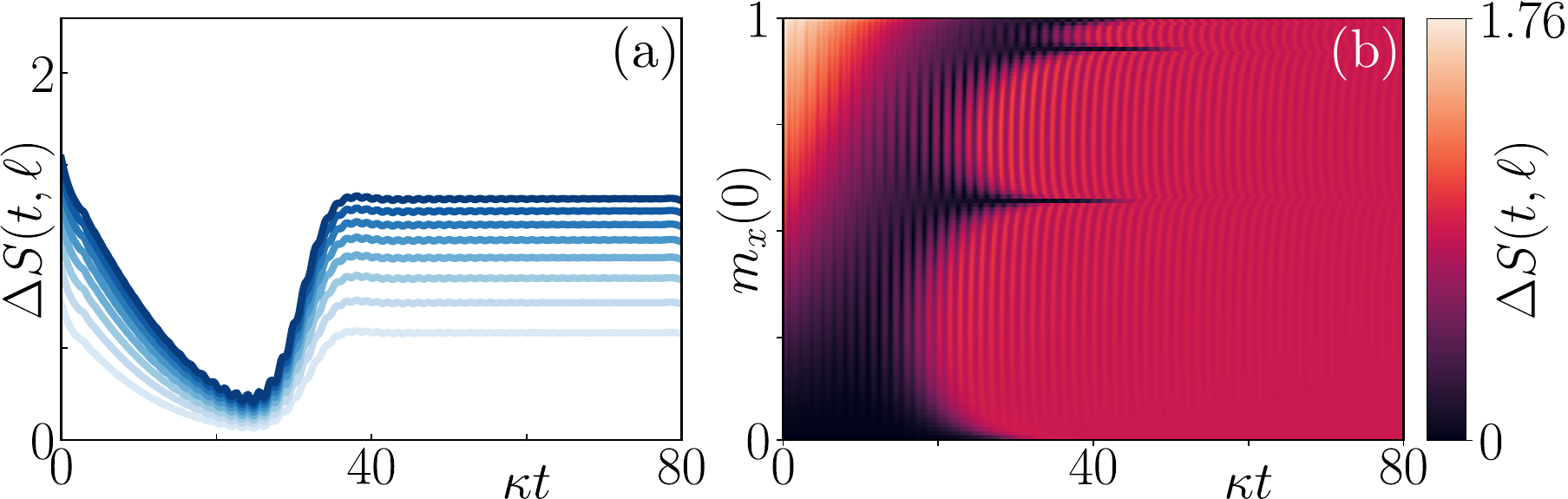}
    \caption{Dynamics of the REA, $\Delta S(t, \ell)$, with respect to the $U(1)$ symmetry in the broken-symmetry phase. We consider a value of $g/\kappa > g_{\rm c}/\kappa$, such that $\Delta S(\infty) = \frac{1}{2}\Delta S ^{\max}_{\ell = 8}$ ($g/\kappa\approx 1.373$). All other parameters are as in Fig.~\ref{fig:stationary_asymm}. (a) REA as a function of time for different subsystem sizes $\ell = 2, 4, 6, \dots, 18$ (going from light to dark). The initial state of the system was chosen to be $(m_x(0), m_y(0), m_z(0)) = (\sqrt{1/3},\: 0, -\sqrt{2/3})$. (b) Density plot of the REA as function of time  and $m_x(0)$ for $\ell = 8$, with $m_y(0) = 0$ and $m_z(0) \leq 0$.}
    \label{U1_broken_symm_zneg}
\end{figure}

In the broken-symmetry phase, see Fig.~\ref{U1_broken_symm_zneg}, the dynamics of the REA is instead non-monotonic. We also see that the stationary state relaxation time is in general longer for states which are initially more asymmetric. The plot shows the same cusp-like features that can be observed in Fig.~\ref{fig:U1_symm_break}(b). 

\end{document}